\documentclass[a4paper,11pt]{article}
\usepackage{pos}
\title{The test and calibration system for the Elementary Cells of the Cherenkov Camera in the PBR Mission}
\ShortTitle{Calibration system for the CC-ECs in the PBR Mission}
\author[a]{Rossella Caruso}
\affiliation[a]{Department of Physics and Astronomy ``E.Majorana'', University of Catania and INFN-Catania,\\
Via Santa Sofia 64, 95123 Catania, Italy}
\onbehalf{on behalf of the JEM-EUSO Collaboration\\[-1mm]{\normalsize \normalfont (a complete list of authors can be found at the end of the proceedings)}}
\emailAdd{rossella.caruso@dfa.unict.it}
\emailAdd{rossella.caruso@ct.infn.it}
\abstract{The development of detectors using Silicon Photo-Multipliers
for acquisition of fast light signals coming from Cherenkov and
fluorescence emissions started by particle showers in the terrestrial
atmosphere is the main goal of the Italian ASI/INFN Agreement
n.2021-8-HH.2-2022, named ``ASI/INFN\_EUSO-SPB2'', in view of the next generation of telescopes in balloon-borne and space-based experiments.  A survey of performances of different Silicon Photo-Multipliers available on the market has been performed to identify the best sensors for space applications, where high thermal excursions and environmental radiation must be mainly taken into account in contrast to ground-based experiments. In particular, a characterization protocol for Silicon Photo-Multiplier qualification has been specified to Hamamatsu S13161-3050AE-08 sensor ($8 \times 8$) array in the $30 \, \textrm{C}^{\circ}$ down to $-40 \, \textrm{C}^{\circ}$ temperature range. The protocol specifies measurements of break-down voltage, quenching resistance, gain, dark count rate and the probability of cross-talk. These parameters have been measured as a function of temperature at fixed over-voltage.  Based on these previous measurements, a dedicated set-up is under completion for performing massive tests to validate and calibrate 32 Silicon Photo-Multipliers  (Hamamatsu S13361-3050 series, 64 channels each), composing the (2048 pixels, $12^{\circ} \times 6^{\circ}$ field of view) Focal Surface of the Cherenkov Camera that will fly on the POEMMA-Balloon with Radio mission.  The technical details and description of this system and the procedural steps implemented are reported, preliminary measurements on the first Elementary Cell are also shown.}

\ConferenceLogo{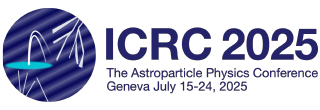}
\FullConference{39th International Cosmic Ray Conference (ICRC2025)\\
 15--24 July 2025\\
Geneva, Switzerland\\}
\begin{document}
\maketitle
\section{Preface}
In the last years, novel semiconductor photo-detectors, such as Silicon Photo-Multipliers (SiPMs) (see Sec.\ref{SIPMs}), have been developed for fundamental science experiments, constituting the enabling technology for a diverse and rapidly growing range of applications: experimental physics, medical imaging and commercial applications are only a few examples. In the recent period, especially, in the framework of the astroparticle physics, an intense investigation of using SiPMs in space applications is ongoing for the next generation of telescopes in balloon-borne and space-based experiments. In the present contribution, a preparatory phase (see Sec.\ref{ASI}), devoted to the development of detectors based on Silicon photo-sensors for acquisition of fast light signals coming from Cherenkov and fluorescence emissions started by charged particles during the development of the Extensive Air Showers (EAS) in the Earth's atmosphere, is introduced. It represents the base for an implementation in Cherenkov and Fluorescence telescopes of the JEM-EUSO Program and, specifically, for the next intermediate sub-orbital mission, named POEMMA Balloon with Radio (PBR) (see Sec.\ref{JEMEUSO}).
\section{Pearls and pitfalls of using SiPMs as photo-detectors in space-based experiments} \label{SIPMs} 
A SiPM is composed of a matrix of identical micro-cells, each one
consisting of a so-called Geiger-mode Avalanche Photo-diode (G-APD) or
Single-Photon Avalanche Diodes (SPAD) and a quenching resistor (R$_q$)
connected in series. The micro-cells are connected in parallel to a bias
voltage (V$_\mathrm{bias}$). SiPMs are also known as Multi-Pixel Photon Counters (MPPCs), naming a micro-cell as pixel. Every pixel produces the same signal when it is hit by a photon and the sum of the pixel responses gives the channel output. The typical dimension of a SiPM sensor is between $(1\times 1) \, \textrm{mm}^2 \div (6 \times 6)  \, \textrm{mm}^2$ and the number of micro-cells per device ranges from several hundreds to several tens of thousands. Micro-cells vary between $(10\times 10) \, \mu\textrm{m}^2 \div (100 \times 100)  \, \mu\textrm{m}^2$ in size, as an example: Hamamatsu S13161-3050AE-08 with channel of $(6 \times 6)  \, \textrm{mm}^2$ and micro-cell $(50 \times 50)  \, \mu\textrm{m}^2$  \cite{SiPMs1}, \cite{SiPMs2}. SiPMs offer numerous advantages over conventional vacuum tube Photo-Multiplier Tubes (PMTs), such as: \\
\textbf {- solid-state technology}: they are solid-state devices, meaning they are more resistant to mechanical shocks, vibrations, and extreme temperatures, which is beneficial for harshest environments; \\
\textbf {- compact size and low voltage operation}: they are significantly smaller and require much lower operating voltages ($10 \div 100$ times lower than PMTs), simplifying electronics and reducing power consumption;\\
\textbf {- cost-effectiveness}: the mass production of Silicon electronics makes them generally less expensive; \\
\textbf {- modularity}: they present an opportunity for constructing modular detectors surfaces due to their compact and lightweight design. To build large sensitive areas, SiPMs can be arranged in arrays; \\
\textbf {- magnetic field immunity}: they are largely unaffected by magnetic fields;\\
\textbf {- high Photon Detection Efficiency (PDE)}: in the red to near-infrared range, they can achieve higher quantum efficiencies than available PMT photocathode materials;\\ 
\textbf {- large Gain (G) and output signal}: they provide a high gain and a large output signal, capable of detecting single photons; \\
\textbf {- low excess noise factor}: they often have a more deterministic photo-electron gain, leading to a lower or negligible excess noise factor and potentially a better signal-to-noise ratio; \\
\textbf {- dynamic range}: by using large arrays of SiPMs, their dynamic range can be significantly expanded, allowing for faster imaging rates or higher signal-to-noise ratios without saturation issues. \\
Nevertheless, SiPMs are usually more sensitive in their performances to the temperature and present also other disadvantages, such as:\\
\textbf {- cost and area}: while generally cheaper for smaller areas, the cost per unit area of SiPMs can be comparable to or even higher than PMTs for larger detection areas;\\
\textbf {- high Dark Count Rate (DCR)}: they exhibit a significantly higher DCR, which can be problematic in low-light applications. DCR is also highly dependent on temperature, requiring cooling for optimal performance; \\
\textbf {- radiation sensitivity}: they are susceptible to radiation damage, particularly from displacement damage caused by high-energy hadrons, leading to an increase in DCR and degradation of performance over time; \\
\textbf {- probability Cross-Talk (pCT)}: the phenomenon where a single photon can trigger a cascade effect that fires multiple micro-cells within the SiPM, leading to increased noise and reduced signal-to-noise ratio; \\
\textbf {- limited dynamic range and band-width}: the limited dynamic range of SiPMs can cause non-linear distortions in the received signal at higher light intensities, and their limited band-width can restrict the maximum transmission rate in certain applications; \\
\textbf {- sensitivity below 400 nm and higher wavelengths}: while some SiPMs have improved near-Ultra-Violet sensitivity, traditional SiPMs are not highly sensitive below 400 nm (where Cherenkov light is emitted) but are too sensitive at higher wavelengths, increasing susceptibility to night sky background noise. \\ 
The design, assembly and completion of a telescope for space applications based on a focal surface formed by arrays of SiPMs must take into account all the above-mentioned factors and achieves a proper balance among pearls and pitfalls of SIPMs. 
\section{The preparatory phase of the ASI/INFN\_EUSO-SPB2 Project} \label{ASI} 
\begin{figure}[h!]
\centering
\includegraphics[width=0.50\textwidth]{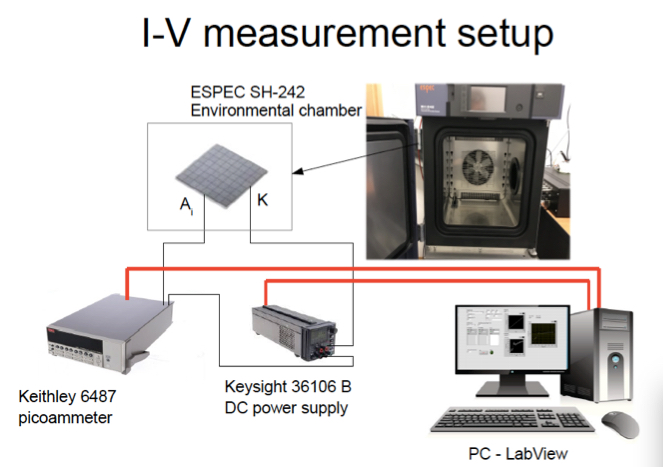}
\hspace{0.3cm}
\includegraphics[width=0.45\textwidth]{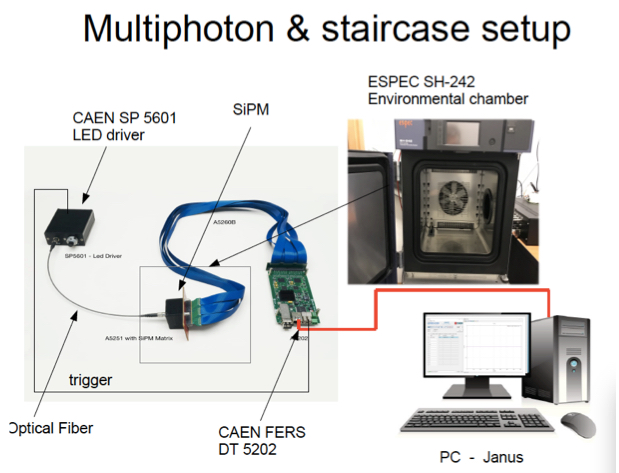}
\caption{\footnotesize{Left: the experimental set-up for measuring the SiPM $(I-V)$ curves at different temperatures. Right: the experimental set-up for measuring the SiPM multi-peak spectrum at different temperatures.}}
\label{set-ups}  
\end{figure}

The development of detectors based on SiPM photosensors  for acquisition
of fast signals coming from Cherenkov and fluorescence emission started
by EASs in the atmosphere, is the main goal of the current ASI/INFN
Agreement n.2021-8-HH.2-2022 and its following amendments and
agreements, signed between the Italian Space Agency (ASI) and the
National Institute of Nuclear Physics (INFN), named ``ASI/INFN\_EUSO-SPB2 (Extreme Universe Space Observatory - Super Pressure Balloon 2)''. It is  devoted to the design, development, advancement, production and completion of a Cherenkov and/or Fluorescence telescope prototype for the next generation of balloon-borne and space-based experiments, with the ultimate purpose of enhancing the Technology Readiness Level (TRL) of such technique.
%and their maturity for flight-qualification, in particular / of mechanics, hardware, front-end, read-out and data acquisition electronics, firmware and software for triggers and slow-control systems, algorithms and codes for simulation, reconstruction and data analysis and also for atmospheric monitoring with assembly and completion of a telescope prototype, including all the test and integration pre-flight campaigns. 
In such framework, the study of performances of different SiPMs available on the market has been performed to identify the best sensors for space applications, where high thermal excursions and environmental radiation must be mainly taken into account with respect to ground-based experiment for indirect measurements of Ultra-High-Energy Cosmic Rays (UHECRs). A specific Working Package (WP), identified as \textit{WP4400:Characterization, selection and validation of SiPMs} has been based on the definition of a procedure in order to characterize and select the SiPM sensors that best fit experimental requirements for space applications. The most important outcome of a long research activity on various SiPMs was a complete study concerning an ($8\times 8$) SiPM matrix (Hamamatsu S13360-3050AE-08) \cite{Hamamatsu}, showing the best performances over a wide range of temperatures by evaluating the Rq, the $V_{\textrm{bd}}$ break-down voltage, the G, the pCT and the DCR. Two different configurations were adopted (see fig.\ref{set-ups}): the first used a pico-ammeter to measure accurately the leakage currents in order to determine the V$_{\textrm{bd}}$ and $R_\textrm{q}$ by means of the I current and V voltages $\textrm{(I-V)}$ curves, both in \textit{forward} and \textit{reverse} modes, and the second one adopted a CAEN DT5202 Front-End and Read-out System (FERS) device to evaluate G, DCR and pCT by means of the multi-peak spectrum. \cite{NIMA-SiPMs}, \cite{ASAP2025}.
 \section{The JEM-EUSO Program and the PBR mission} \label{JEMEUSO} 
Since 2010, the international JEM-EUSO (Joint Exploratory Missions for Extreme Universe Space Observatory) Collaboration (URL: \href{https://www.jemeuso.org/} {\texttt{https://www.jemeuso.org}}), around 170 researchers, 60 institutions and 16 countries has been developing an ambitious Program (\cite{JEM-EUSO1}, \cite{JEM-EUSO2}) with the support of major International and National Space Agencies (ASI, CNES, ESA, JAXA, NASA, ROSCOMOS) and research funding institutions, to enable UHECR (E>100 PeV) and Very High-Energy (VHE) neutrino (E>1 PeV) observations from space. Its main objective is to develop a large mission with dedicated instrumentation looking down on the Earth's atmosphere from space, both towards nadir and/or towards the limb direction, to detect the EAS initiated by such particles in the high atmosphere. This strategy is intended to complement the observations made with ground-based observatories by allowing a significant increase in the exposure and by achieving near-uniform full sky coverage. In the last decade, the JEM-EUSO Collaboration effectively developed five intermediate missions: one ground based (EUSO-TA \cite{EUSO-TA}), three balloon-borne (EUSO-Balloon \cite{EUSO-Balloon}, EUSO-SPB1 \cite{SPB1}, EUSO-SPB2 \cite{SPB2} and one space-based (Mini-EUSO \cite{MiniEUSO},\cite{MiniEUSO2}). Recently, a new balloon-borne flight has been approved and funded by NASA for a launch planned in Spring 2027 from the Wanaka base in New Zealand. This next intermediate mission, PBR (\cite{PBR}), will be the closest prototype to date for the large-size space-based POEMMA (Probe Of Extreme Multi-Messenger Astrophysics) mission \cite{POEMMA}, a stereo double telescope to be considered for a NASA Probe Mission in the next decade. The PBR Mission is a scientific mission involving an instrument designed to be carried by a NASA suborbital Super Pressure Balloon (SPB) with a projected duration of up to 100 days, circling over the Southern Ocean. The PBR instrument consists of a 1.1 m aperture Schmidt telescope \cite{Optics}, similar to the POEMMA, with two photo-detectors in its hybrid Focal Surface (FS): a Fluorescence Camera (FC) \cite{FC} and a Cherenkov Camera (CC) \cite{CC}, both mounted on a frame that can be tilted \cite{Tilt} to point from nadir up to 13 degrees above the horizon. In addition, PBR has a Radio Instrument (RI) optimized for the detection of EASs, covering the $(50 \div 550)$ MHz range. 
\section{The calibration system of the Elementary Cell units and first results} 
\begin{figure}[h!]
%\centering
\includegraphics[width=0.30\textwidth]{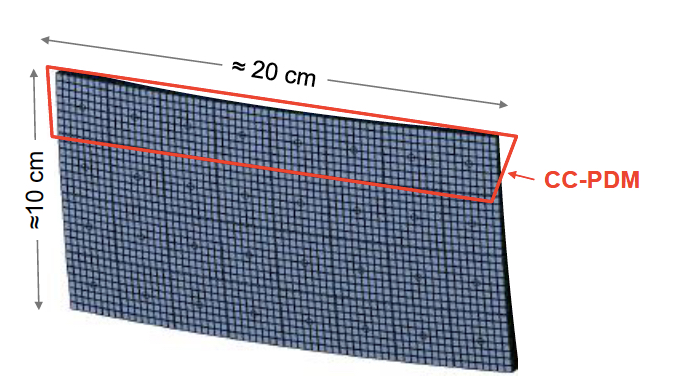}
\includegraphics[width=0.30\textwidth]{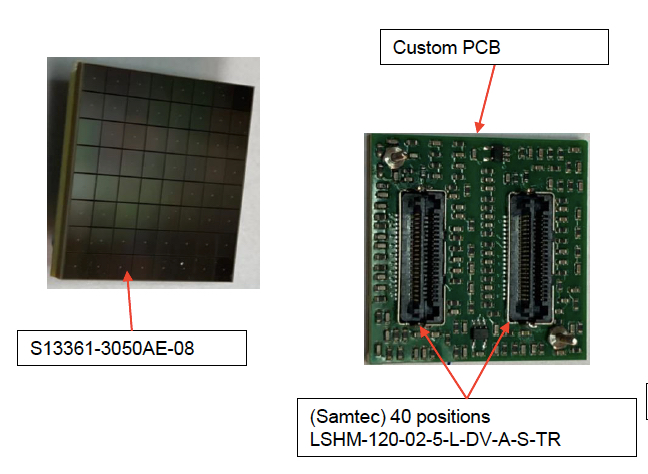}
%\hspace{0.3cm}
\includegraphics[width=0.30\textwidth]{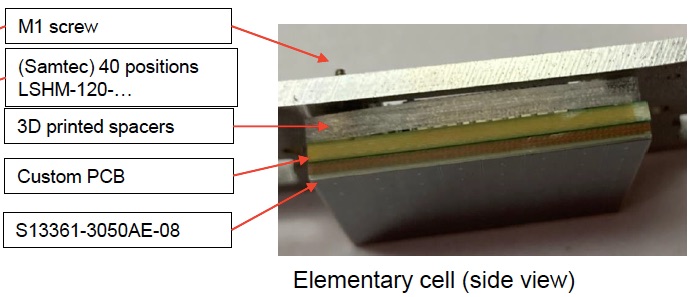}
\caption{\footnotesize{From left to right: an artistic view of the Focal Surface; different components of an Elementary Cell.}}
\label{EC}  
\end{figure}

\begin{figure}[h!]
\centering
\includegraphics[width=0.70\textwidth]{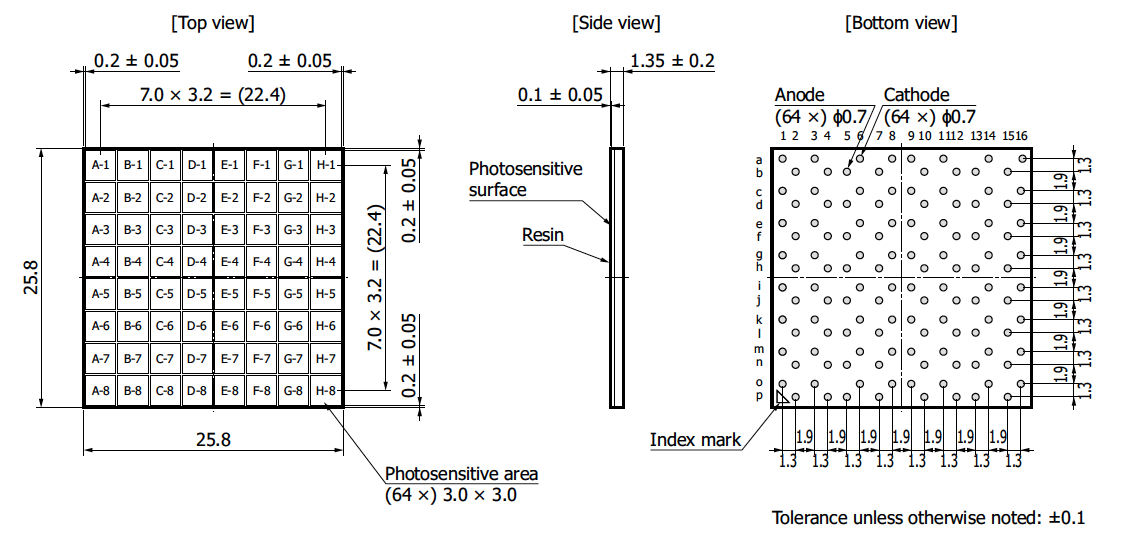}
\includegraphics[width=0.30\textwidth]{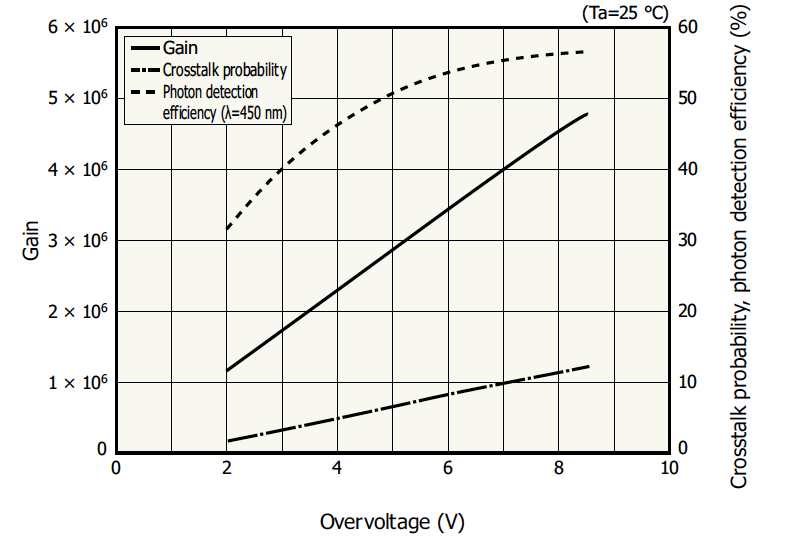}
\hspace{0.3cm}
\includegraphics[width=0.60\textwidth]{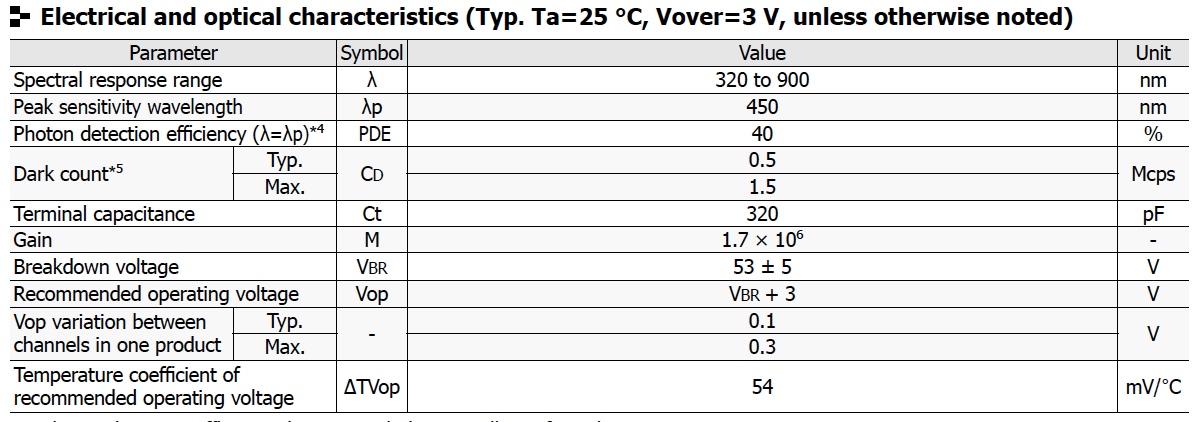}
\caption{\footnotesize{Above: matrix scheme and geometrical dimensions of the Hamamatsu SiPM S13161-3050AE-08 model (top, side and bottom views). Below, on the left: gain, probability of cross-talk and photon-detection efficiency as function of the over-voltage at room temperature; on the right: a table reporting electrical and optical characteristics.}}
\label{S13}  
\end{figure}

\begin{figure}[h!]
\centering
\includegraphics[width=0.45\textwidth]{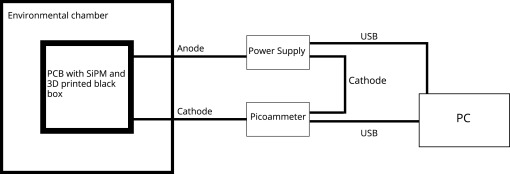}
\hspace{0.5cm}
\includegraphics[width=0.45\textwidth]{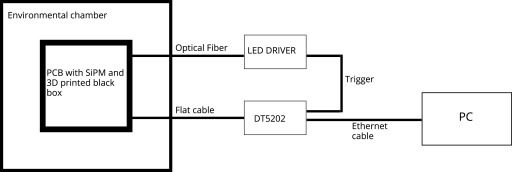}
\caption{\footnotesize{Left: the block diagram of the experimental set-up used for measuring the SiPM $(I-V)$ curves at different temperatures. Right: the block diagram of the experimental set-up used for measuring the SiPM multi-peak spectra at different temperatures.}}
\label{ECsetup}  
\end{figure}

\begin{figure}[h!]
\centering
\includegraphics[width=0.25\textwidth]{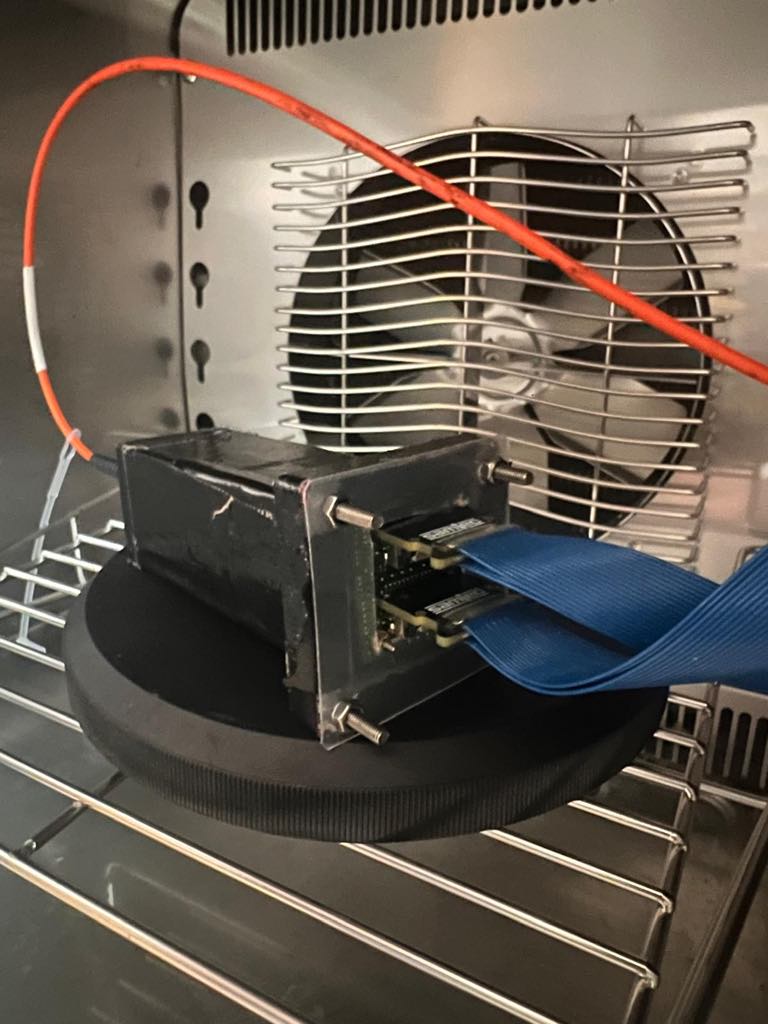}
\hspace{0.2cm}
\includegraphics[width=0.35\textwidth]{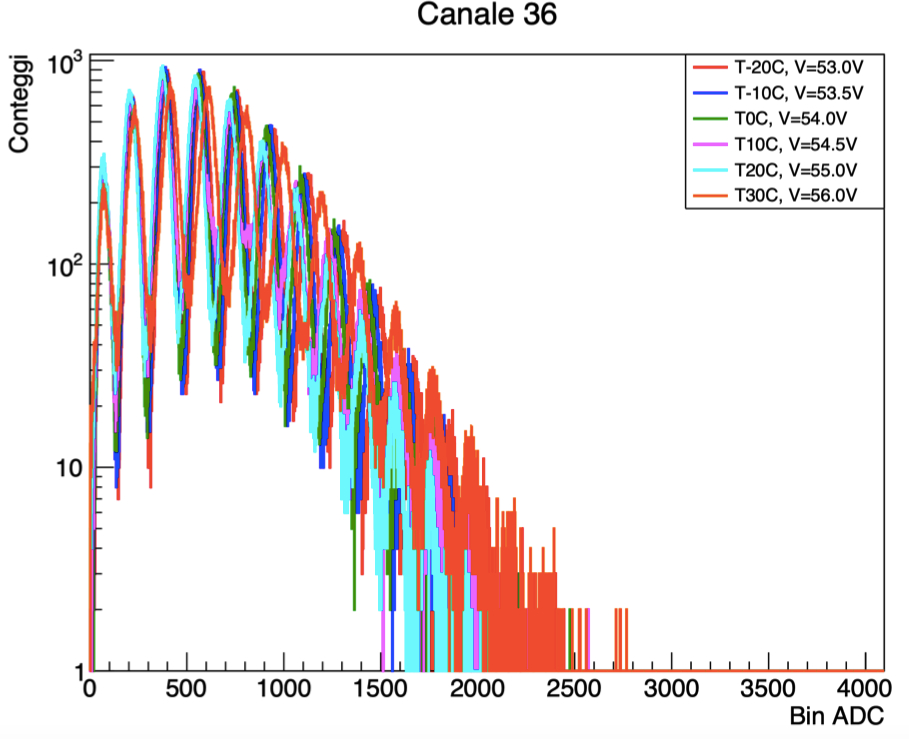}
\hspace{0.2cm}
\includegraphics[width=0.35\textwidth]{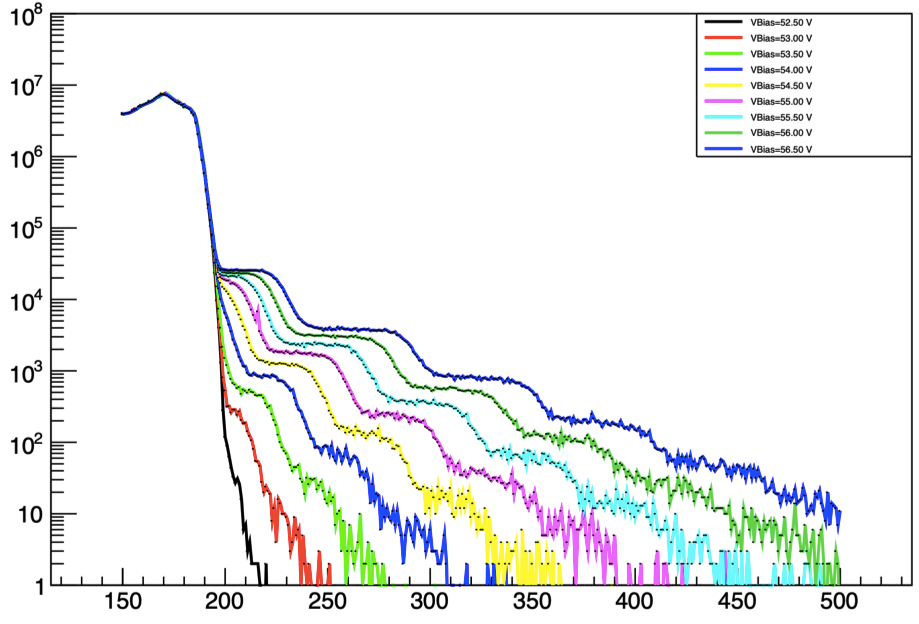}
\caption{\footnotesize{Left: the Elementary Cell inside the climate chamber joined to a custom 3D-printed light-tight box, placed on a designated base, illuminated by means of an optical fiber (in red) and connected with the instrumentation outside through Samtec cables (in blue); centre: multi-peak spectrum (at LED on) measured for a given channel of the first EC at fixed V$_{bias}$ at different temperatures; right: dark count spectrum (at LED off) measured for the same given channel, at fixed temperature ($ 25^{\circ} \,\textrm{C}$) for different V$_{bias}$.}}
\label{results}  
\end{figure}

The Focal Surface (see fig.\ref{EC}) of the PBR Camera Cherenkov (CC)
contains four rows of 8 Elementary Cells (ECs). Each EC is a ``sandwich'' (see fig.\ref{EC}) composed by a SiPM ($8 \times 8$ channels) matrix, a custom Printed Circuit Board (PCB), supporting two LSHM-120-02.5-L-DV-A-S-TR Samtec high-speed connectors (2 rows, 40 positions) and a 3D printed spacer, joined by means of M1 screws. Two (40 positions) HLCD-20-12.00-TD-TH1 Samtec 38 AWG micro-coax cables  connect the EC to the Front-End and Read-out electronics.
%\begin{figure}[h!]
%\centering
%\includegraphics[width=0.70\textwidth]{fig/FS.jpg}
%\caption{\footnotesize{An artistic view about the Focal Surface of the PBR Cherenkov Camera.}}
%\label{FS}  
%\end{figure}
Each SiPMs matrix is an Hamamatsu (S13161-3050AE-08 model) photo-sensor, with specific properties reported in fig.\ref{S13}, consisting of 64 channels of $(3\times3$) mm$^2$ pixels, totaling 32 SiPM matrixes and 2048 pixels for the entire camera. 
Each row (8 ECs, 512 pixels) constitute a CC Photo Detection Module (or CC-PDM). These CC-PDMs are mounted on a structure that shapes the photo-sensitive plane to approximate the spherical curvature required by the telescope optics. The total FoV spans $(12^{\circ} \times 6^{\circ})$, with each pixel covering 0.2$^{\circ}$. The sensitive wavelength range is $(320 \div 900)$ nm. The Read-Out electronics will be either based on RadioROC or MIZAR ASICs \cite{MIZAR}, which are under development by providing a sampling frequency of 200 MHz.
\\
A characterization protocol for the EC qualification in the $30 \, \textrm{C}^{\circ}$ down to $  -40 \, \textrm{C}^{\circ}$ temperature range has been specified . The protocol specifies measurements of $V_{\textrm{bd}}$, $R_\textrm{q}$, G, DCR and the pCT. These key parameters can been measured as a function of temperature at fixed V$_{\textrm{ov}}$ for each channel of each EC. A dedicated experimental system - composed by two different set-ups, one for SiPMs measurements of $(I-V)$ curves and the other for SiPMs measurements of multi-peak spectra  (see fig.\ref{ECsetup}) - has been recently finalized for performing massive tests to test and calibrate the 32 Electronic Cells, one at a time. Both set-ups make use of an SH-242 environmental chamber produced by ESPEC  hosting the EC, to perform tests at stable and different temperatures. This chamber spans $(-40 \div +150) \pm 0.5^{\circ} \, \textrm{C}$  temperature range; its interior size is $(300 \times 250 \times 300) \, \textrm{mm}^3$. In order to connect the EC to the instrumentations, located outside the climatic cell, 105782-120-DH-S Samtec flat cables (1 m length)  are used. The EC is mounted on purpose inside a custom 3D-printed light-tight box and properly illuminated at 402 nm by a LED driver (CAEN SP5601 model), fed through a 40 cm CAEN AI2740 FC optical fiber interfaced to the box. One set-up (on the left in fig. \ref{ECsetup}) allows powering individual anodes and collect the leakage current from the cathode in the \textit{reverse} measurement, and vice-versa in the \textit{forward} one.  A Keysight e36106b DC power supply [was used to power the EC while a Keithley 6487 pico-ammeter has been adopted for precise current measurements. Both devices were automatically operated using a specifically developed LabView program that allows setting the voltage range for supplying the EC, the stabilization time, used for stabilizing the EC after each voltage change, and the measurement time, the time interval during the current is acquired. In the second set-up (on the right, in fig. \ref{results}), the EC is directly connected to the CAEN DT5202 FERS module based on 2 Citiroc-1 A chips produced by Weeroc. The module permits carrying out photon counting, saving the signal Time Stamp and the Time over Threshold for each signal, accepting an external trigger (in such case, from the LED driver). The data acquisition is automatically operated using Janus software, an open source software specifically developed for the control and readout of FERS-5200 boards. The data archive and sharing is operated by using Notion software, an open AI (Artificial Intelligence) Workspace for archive and sharing data files, documents, plots, etc. In order to characterize the most key parameters of the ECs, all the measurements were performed at different temperatures for all 64 channels, ranging in $(-40 \div +30)^{\circ} \,\textrm{C}$ with a step of $ 10^{\circ} \,\textrm{C}$ and at $ 25^{\circ} \,\textrm{C}$ with different V$_{\textrm{ov}}$ . The EC characterization started evaluating V$_{\textrm{ov}}$  and $R_\textrm{q}$ and fixing an operating voltage for each temperature, then G, DCR and pCT are evaluated as a function of temperature at fixed V$_{\textrm{ov}}$. A complete procedure (i.e acquisition of $(I-V)$ curves and multi-peak spectra for all the channels of a single SiPM matrix, under thermal excursions in the fixed range) for each EC require a long duration - about ($ 10 \div16$)  hours - mainly due to  dead-times by the climate chamber needed for ranging to and from higher and lower temperatures. A preliminary full measurement on the first EC, in order to test the instrumentation, to validate the procedure, to fix the key parameters, to optimize the data acquisition softwares and to implement the algorithm codes for data analysis, has been successfully fulfilled (as an example, preliminary results are reported in fig.\ref{results}).  
%\begin{figure}[h!]
%\centering
%\includegraphics[width=0.20\textwidth]{fig/ECsetup_pic1.jpeg}
%\hspace{0.5cm}
%\includegraphics[width=0.25\textwidth]{fig/map.jpg}
%\caption{\footnotesize{Left: the Elementary Cell inside the climate chamber above joined to a custom 3D-printed light-tight box, placed on a designated base, illuminated by means of an optical fiber (in red) and connected with the instrumentation outside through Samtec cables (in blue). Right: light intensity map of the SiPM matrix.}}
%\label{ECsetup}  
%\end{figure}
\\
For the time being, a fine-tuning of the whole test and calibration system is in progress, algorithm codes for analysing data are under revision, automation of the whole procedure for performing EC massive characterization is under study and the optimization of timings, based on the selection of a limited number of channels representative of the SiPM matrix, is taken into consideration.  
%Moreover, in the meantime, an upgrade of the System - including an integrating sphere and an $XYZ$ translator plane - devoted to the measurement of EC PDE and uniformity is also in preparation.

\acknowledgments
\small{The authors would like to acknowledge the support by NASA award 80NSSC22K1488 and 80NSSC24K1\\780, by the French space agency CNES and the
Italian Space agency ASI. The work is supported by OP JAC financed by
ESIF and the MEYS CZ.02.01.01/00/22$\_$008/0004596. We gratefully
acknowledge the collaboration and expert advice provided by the PUEO
collaboration. This research used resources of the National Energy
Research Scientific Computing Center (NERSC), a U.S. Department of
Energy Office of Science User Facility operated under Contract
No.DE-AC02-05CH11231. We acknowledge the NASA Balloon Program Office and
the Columbia Scientific Balloon Facility and staff for support. We also
acknowledge the invaluable contributions of the administrative and
technical staffs at our home institutions.  In particular, this work
would not have been possible without the financial support of the
ASI-INFN agreement n. 2021-8-HH.0, its amendments and its and agreement
n. 2020-26-Hh.0., Research Project ``EUSO-SPB2 (Extreme Universe Space
ObservatorySuper Pressure Balloon)'', WP4400 ``Characterization, Selection and Test of SiPM tiles''.}

\newpage
%\newpage
%{\Large\bf Full Authors list: The JEM-EUSO Collaboration}
%{\scriptsize (author-list as of September 15th, 2025 )} \hspace{0.6cm}
%{\scriptsize (version  \today{} \currenttime{})}
%\vspace*{0.5cm}
%contact: zbigniew.plebaniak@roma2.infn.it, marco.ricci@lnf.infn.it
{\Large\bf Full Authors list: The JEM-EUSO Collaboration}
%{\scriptsize (author-list as of August 20th, 2025 with reorganized affiliations)} \hspace{0.6cm}
%{\scriptsize (version  \today{} \currenttime{})}
%\vspace*{0.5cm}
%contact: zbigniew.plebaniak@roma2.infn.it, marco.ricci@lnf.infn.it

\begin{sloppypar}
{\small \noindent
M.~Abdullahi$^{ep,er}$              % Italy
M.~Abrate$^{ek,el}$,                % Italy
J.H.~Adams Jr.$^{ld}$,              % USA 
D.~Allard$^{cb}$,                   % France
P.~Alldredge$^{ld}$,                % USA
R.~Aloisio$^{ep,er}$,               % Italy
R.~Ammendola$^{ei}$,                % Italy
A.~Anastasio$^{ef}$,                % Italy %%
L.~Anchordoqui$^{le}$,              % USA
V.~Andreoli$^{ek,el}$
% ...
}
\end{sloppypar}
\vspace*{.3cm}

%%\newpage
{ \footnotesize
\noindent
%
% Czech Republic - 2 institutions
%Czech Republic\\
% ...
$^{cb}$ Universit\'e de Paris, AstroParticule et Cosmologie (CNRS), Paris, France\\
% ...
$^{ef}$ Istituto Nazionale di Fisica Nucleare (INFN), Sezione di Napoli, Naples, Italy\\
% ...
$^{ei}$ Istituto Nazionale di Fisica Nucleare (INFN), Sezione di Roma Tor Vergata, Rome, Italy\\
% ...
$^{ek}$ Istituto Nazionale di Fisica Nucleare (INFN), Sezione di Torino, Turin, Italy\\
$^{el}$ Universit\`a di Torino, Dipartimento di Fisica, Turin, Italy\\
% ...
$^{ep}$ Gran Sasso Science Institute (GSSI), L’Aquila, Italy\\
$^{er}$ Istituto Nazionale di Fisica Nucleare (INFN), Laboratori Nazionali del Gran Sasso, Assergi, Italy\\
% ...
$^{ld}$ University of Alabama in Huntsville, Huntsville, AL, USA\\
$^{le}$ City University of New York (CUNY), Lehman College, Bronx, NY, USA\\
% ...
}

\end{document}